\documentclass[pra,aps,onecolumn,nopacs,superscriptaddress,nofootinbib]{revtex4}
\usepackage[T1]{fontenc}
\usepackage[latin1]{inputenc}
\usepackage{lmodern}\usepackage{graphicx}
\usepackage{dcolumn}
\usepackage{bm}
\usepackage{amsmath}
\usepackage{amssymb}
\usepackage{pst-all}
\usepackage{psfrag}
\usepackage{epsfig}
\usepackage{units}

  \def\ee{{\rm e}}   
      
    \def\Eb{{\bf E}}  \def\rb{{\bf r}}

\def\EF{{E_{\text{F}}}}

\begin{document}

\title{Electrical Detection of Single Graphene Plasmons}

\author{Renwen Yu}
\affiliation{ICFO-Institut de Ciencies Fotoniques, The Barcelona Institute of Science and Technology, 08860 Castelldefels (Barcelona), Spain}
\author{F. Javier Garc\'{\i}a de Abajo}
\affiliation{ICFO-Institut de Ciencies Fotoniques, The Barcelona Institute of Science and Technology, 08860 Castelldefels (Barcelona), Spain}
\affiliation{ICREA-Instituci\'o Catalana de Recerca i Estudis Avan\c{c}ats, Passeig Llu\'{\i}s Companys 23, 08010 Barcelona, Spain}
\email{javier.garciadeabajo@icfo.es}

\begin{abstract}
Plasmons --the collective oscillations of electrons in conducting materials-- play a pivotal role in nanophotonics because of their ability to couple electronic and photonic degrees of freedom. In particular, plasmons in graphene --the atomically thin carbon material-- offer strong spatial confinement and long lifetimes, accompanied by extraordinary optoelectronic properties derived from its peculiar electronic band structure. Understandably, this material has generated great expectations for its application to enhanced integrated devices. However, an efficient scheme for detecting graphene plasmons remains a challenge. Here we show that extremely compact graphene nanostructures are capable of realizing on-chip electrical detection of single plasmons. Specifically, we predict a twofold increase in the electrical current across a graphene nanostructure junction caused by the excitation of a single plasmon. This effect, which is due to the increase in electron temperature following plasmon decay, should persist during a picosecond time interval characteristic of electron-gas relaxation. We further show that a broad spectral detection range is accessible either by electrically doping the junction or by varying the size of the nanostructure. The proposed graphene plasmometer could find application as a basic component of future optics-free integrated nanoplasmonic devices.
\end{abstract}
\maketitle

\section{Introduction}

Plasmons have the ability of focusing light down to nanometer-sized regions, where the optical field intensity is amplified by orders of magnitude \cite{LSB03}. These properties have been extensively used to develop detection techniques with a sensitivity down to the single molecule level \cite{KWK97,NE97,XBK99,M05_2,paper125}, trigger efficient nonlinear processes at the nanoscale \cite{DN07,PN08}, enhance optoelectronic devices for light harvesting \cite{CP08,AP10}, spectrometry \cite{KSN11,LCL11,CSB14}, and photocatalisis \cite{SLL12_2,GZT12,MLL13,MLS13,C14,MZG14}, and assist tumor diagnosis \cite{QPA08} and treatment \cite{HSB03}, among other feasts. However, traditional plasmonic materials such as noble metals present a relatively large level of ohmic losses, which limit the plasmon lifetime to tens of femtoseconds in metallic nanostructures \cite{JC1972,paper242}. Additionally, the extreme optical confinement comes at a prize, as it further contributes to limit the interaction with external light. As a result, typical light-to-plasmon coupling cross-sections are barely exceeding the projected geometrical area of the plasmon-supporting structures \cite{paper242}. In an attempt to circumvent this problem, electrical generation and detection of plasmons has been pursued to substitute optics-based methods \cite{FKC09,HDK09_2,GCB15}, although only modest efficiencies have been achieved so far using noble metals.

Graphene has recently emerged as an excellent plasmonic material \cite{WSS06,HD07,JBS09,JGH11,FAB11,SKK11,paper196,FRA12,YLC12,YLL12,paper212,BJS13,YLZ13,paper230,YAL14}, which combines large field confinement and enhancement \cite{paper216} with relatively low losses \cite{WLG15} and the ability to tune the plasmons electrically \cite{JGH11,FAB11,SKK11,paper196,FRA12,YLC12,paper212,BJS13,YLZ13,paper230} or magnetically \cite{YLL12}. In particular, electrical doping using gates can change the Fermi energy of the carbon layer by as much as $\EF\sim1\,$eV from its undoped state \cite{CPB11}, thus opening a $2\EF$ optical gap and effectively sustaining well-behaved plasmons up to energies $\sim\EF$. A unique characteristic of plasmons in this material is that they are sustained by a comparatively small number of electrons \cite{paper235}. We thus expect that the excitation of a plasmon in a graphene nanostructure will significantly modify the population of the electronic levels, to the extent that its electrical properties will be strongly affected. Plasmons quickly decay into hot electrons, which lead to observable photocurrents \cite{XML09,EBJ11,TPM15} and eventually thermalize to an elevated electron temperature in extended graphene \cite{GSM11,TSJ13,MTP15}

Here we show that the excitation of a single plasmon in a graphene nanostructure produces profound modifications in its electrical properties, which we then use to detect the presence of the plasmon. Our quantum-mechanical calculations confirm the excellent performance of graphene quantum dots \cite{BBN14} for on-chip electrical detection of single plasmons. More precisely, we find a twofold increase in the electrical current passing across a nanographene junction when one of its plasmons is excited. Our predictive simulations are based on a state-of-the-art atomistic quantum-mechanical model incorporating tight-binding states for the conduction electrons \cite{W1947,CGP09}, the random-phase approximation for the linear optical response of structured graphene \cite{paper183}, and the Landauer formalism for the electrical transport properties \cite{L1957,D97,V08}. We consider realistic levels of doping and graphene quality, with an intrinsic electron lifetime $\tau=$66\,fs ({\it i.e.}, $\hbar\tau^{-1}=10\,\mathrm{meV}$), somewhere in between high-quality graphene \cite{WLG15} and small polycyclic aromatic hydrocarbons \cite{paper260}. We remark that the electrical tunability of graphene allows us to shift the plasmon frequency, thus supporting the great potential of our detection scheme for nanoscale spectrometry. 


\begin{figure}
\begin{centering}
\includegraphics[width=1\textwidth]{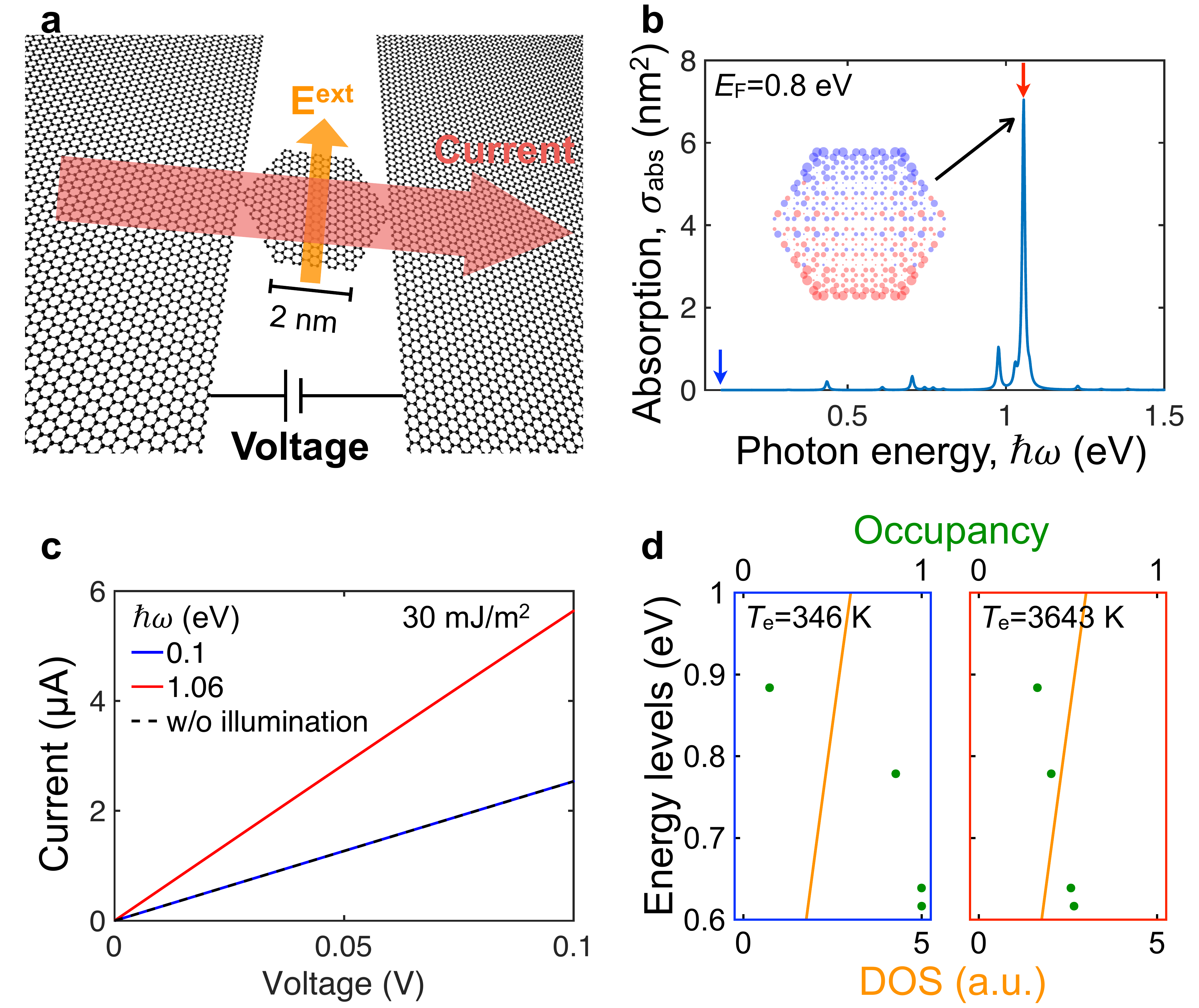}
\par\end{centering}
\caption{{\bf Electrical detection of plasmons in a graphene hexagone quantum dot (GHQD) nanojunction.} ({\bf a}) We show a sketch of the structure under consideration, consisting of a GHQD (2\,nm side length) connected to two semi-infinite graphene sheets that operate as electrical contacts. Currents are measured under an applied DC bias across the nanojunction, possibly in the presence of plasmons excited by external light that is polarized along the perpendicular direction. ({\bf b}) Calculated absorption cross-section of a single GHQD doped to a Fermi energy $\EF=0.8\,$eV. A pronounced plasmon resonance peak is observed at $\sim1\,$eV photon energy, whose associated induced charge in shown as an inset. ({\bf c}) Calculated current as a function of bias voltage in the presence of external illumination (30\,mJ$/$m$^{2}$ light fluence) for two different photon energies, corresponding to on- and off-resonance conditions relative to the prominent plasmon shown in {\bf b} (red and blue curves, respectively). The current in the absence of external illumination is shown as a dashed curve, nearly overlapping with the current obtained under off-resonance illumination. ({\bf d}) Electron energy levels (left scale, horizontal lines) of the GHQD and their occupancies (upper scale and green dots) in the presence of external illlumination under on- and off-resonance conditions (right and left panels, respectively). We also show the density of states (DOS, orange curves, lower scale) of extended graphene and the electron temperature $T_{\rm e}$ right after irradiation (labels). The occupancy is dramatically reduced near the Fermi level (0.8\,eV) under resonant illumination (right panel).}
\label{Fig1}
\end{figure}

\section{Results}

Our proposed structure is based on the concept of molecular junctions \cite{TDH98,TGW01,GN06,BMO02,NR03}, as illustrated in Fig.\ \ref{Fig1}a, where a graphene hexagone quantum dot (GHQD) is contacted to two semi-infinite graphene sheets that act as electrodes. The GHQD (2\,nm side length) is connected to these electrodes through two carbon-carbon bonds on either side, thus configuring a nanojunction. The entire structure is assumed to be doped to a Fermi level of 0.8\,eV. We consider external illumination with polarization perpendicular to the junction. For this polarizaton, the plasmons excited in the GHQD should have the bulk of their induced charge piled up near the sides of the hexagon that are far from the semi-infinite contacts. Therefore, these plasmons are indeed similar to those of the isolated hexagon ({\it i.e.}, without the contact), as observed in the trend of numerical simulations for finite contacts of increasing size (see Fig.\ \ref{Fig6} in the Appendix). Consequently, we neglect the contacts in the optical response and focus on the central part of the nanojunction in order to simplify our analysis. A pronounced plasmon resonance is observed at $\sim1\,$eV in the simulated absorption spectrum (Fig.\ \ref{Fig1}b), which has a dipolar nature (see inset). Importantly, the spectral position of this plasmon resonance is found to be robust against a variation of $0-0.1\,$V in the applied DC bias across the nanojunction (see Fig.\ \ref{Fig7} in the Appendix). 

We now calculate the electrical \textit{I}-\textit{V} response of the nanojunction for different incident photon energies using the Landauer formalism \cite{L1957,D97,V08} (see Appendix for more details). In particular, we show the current right after light pulse irradiation and subsequent thermalization of the electron gas at an elevated temperature $T_{\rm e}$ (see Fig.\ \ref{Te} in the Appendix). The results are shown in Fig.\ \ref{Fig1}c. When the external illumination (30\,mJ$/$m$^{2}$ fluence) is resonant with the dominant GHQD plasmon observed in Fig.\ \ref{Fig1}b, a clear amplification of the current $i_{\mathrm{on}}$ is observed (red curve) compare with the current $i_{\mathrm{off}}$ for off-resonance illumination (blue curve) or the current $i_{0}$ in the absence of external light (dashed curve). The maximum variation predicted by this figure between on- and off-resonance conditions corresponds to a factor of 2 in the observed current.

As expected, the current scales linearly with the DC applied voltage (Fig.\ \ref{Fig1}c). The slope of these curves is related to the number of available electronic levels contained in the central GHQD for energies lying in between the Fermi levels of the two gates, as shown in Fig.\ \ref{Fig1}d. Actually, we can obtain further insight into the plasmon-enhanced conductivity of the nanojunction by examining the distribution of GHQD energy levels with and without excited plasmons (Fig.\ \ref{Fig1}d). In the absence of external illumination or under off-resonance conditions (Fig.\ \ref{Fig1}d, left panel), the island shows several unoccupied electronic states right above the Fermi level. When switching to on-resonance illumination, the occupancy of these energy levels is dramatically modified (Fig.\ \ref{Fig1}d, right panel, green dots) as a result of energy absorption from the optical pulse. This effect is more pronounced for states that are closer to the unbiased Fermi energy, which therefore contribute more efficiently to modify the electrical conductivity of the GHQD, leading to amplification of the currents when plasmons are excited. Another element that controls the obtained current is the electronic density of states (DOS) of the graphene contacts (Fig.\ \ref{Fig1}d, lower scales), which evolves rather smoothly over the relevant range of electron energies (see Appendix and Fig.\ \ref{FigS1}a for more details); as expected, the current decreases when we artificially reduce the DOS (see Fig.\ \ref{FigS1}b). Additionally, the magnitude and sign of the observed effects strongly depend on the graphene Fermi energy (see below).

\begin{figure}
\begin{centering}
\includegraphics[width=1\textwidth]{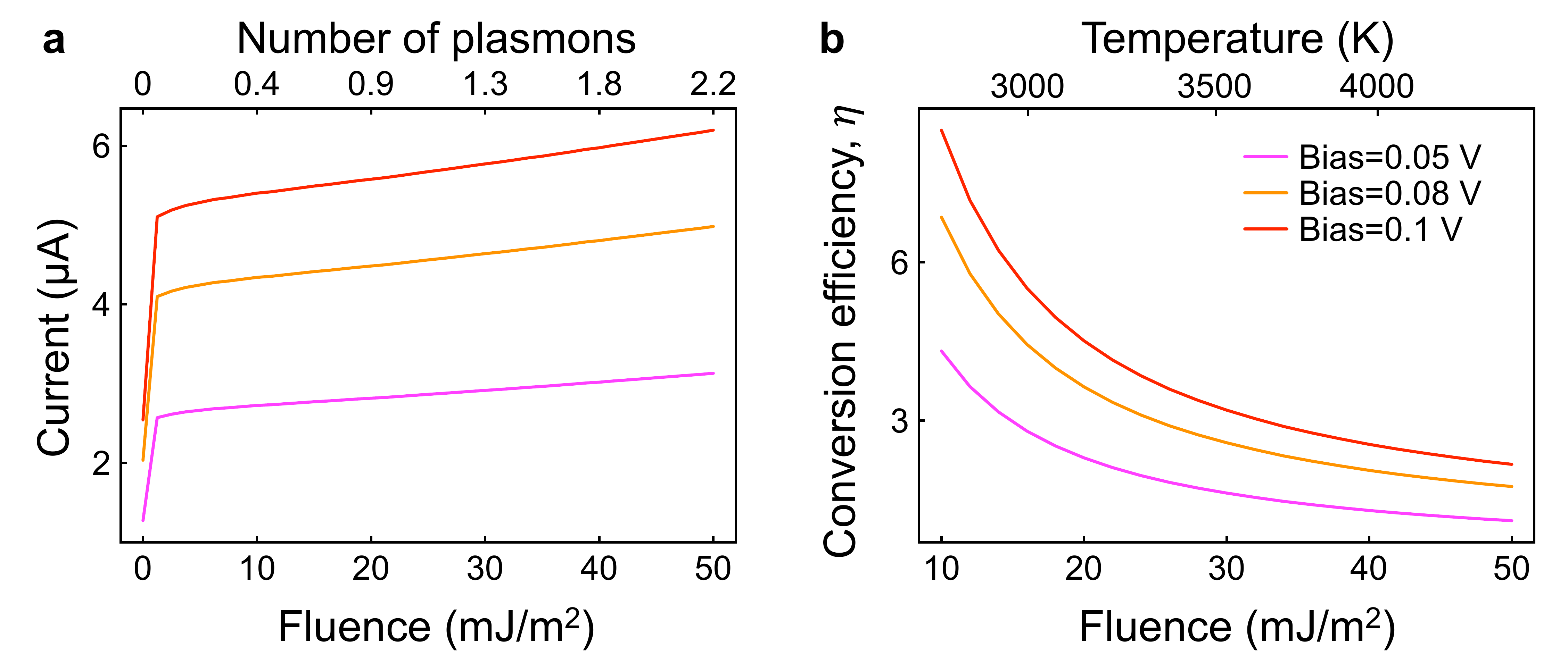}
\par\end{centering}
\caption{{\bf Toward single plasmon detection.} ({\bf a}) Calculated current as a function of incident light fluence (lower scale) for different DC bias voltages (see legend in {\bf a}) in the same structure as in Fig.\ \ref{Fig1}a. We consider a resonant photon energy $\hbar\omega=1.06\,\mathrm{eV}$. ({\bf b}) Plasmon-to-charge conversion efficiency as a function of incident light fluence (see Appendix). The fluence-dependent number of excited plasmons and the electron temperature right after laser pulse irradiation are shown as upper scales in {\bf a} and {\bf b}, respectively (see main text).}
\label{Fig2}
\end{figure}

The number of plasmons that are required to be excited in the GHQD in order to observe a detectable variation in the current across the junction is an important parameter that permits us to assess the performance of the device. We show in Fig.\ \ref{Fig2}a the dependence of the current on light fluence under on-resonance illumination conditions ($\hbar\omega=1.06\,\mathrm{eV}$) for different DC bias voltages applied across the nanojunction. All curves display a similar (weak) linear increase of the current with light fluence (above $\sim1\,$mJ$/$m$^2$), although larger voltages produce more intense currents and reveal a nonlinear dependence of the elevated electron temperature (see Appendix). Incidentally, a depletion of the current takes place in the low-fluence region (see Figs.\ \ref{Fig2}a and \ref{FigS2}), with a characteristic transition temperature $T_{\rm e}\sim200\,$K, which is compatible with the presence of an electronic state at an energy $k_{\rm B}T_{\rm}\sim20\,$meV below the Fermi level (see Fig.\ \ref{Fig1}d). We also show the average number of plasmons sustained by the GHQD (upper scale), as estimated from the optical absorption cross-section, and the plasmon energy (see Appendix). These results indicate that the device is capable of detecting single-plasmons. Nevertheless, the current shows a monotonic increase with light fluence above $\sim1\,$mJ$/$m$^2$, which should directly permit correlating the readout of an amperemeter with the number of plasmons excited by the optical pulse in the graphene island. We thus propose to exploit the graphene nanojunction as a plasmometer with a sensitivity down to the single-plasmon level.

An important figure of merit for our plasmometer is the so-called plasmon-to-charge conversion efficiency $\eta$ \cite{FKC09,GCB15}, which gives the number of additional electrons that circulate through the junction as a result of the excitation of one plasmon (see Appendix). As an conservative estimate of $\eta$, we assume that the electron stays at the initial elevated temperature during $200\,$fs, which is of the order of what is observed in pump-probe experiments \cite{DSC08,WOP11}. Remarkably, the efficiency reaches values of 8 electrons per plasmon (Fig.\ \ref{Fig2}b), which is around 18 times larger than previously obtained results for plasmons propagating along silver nanowires \cite{GCB15}. Additionally, a negative correlation between efficiency and light fluence is observed, consistent with previous experimental studies \cite{GCB15}.

\begin{figure}
\begin{centering}
\includegraphics[width=1\textwidth]{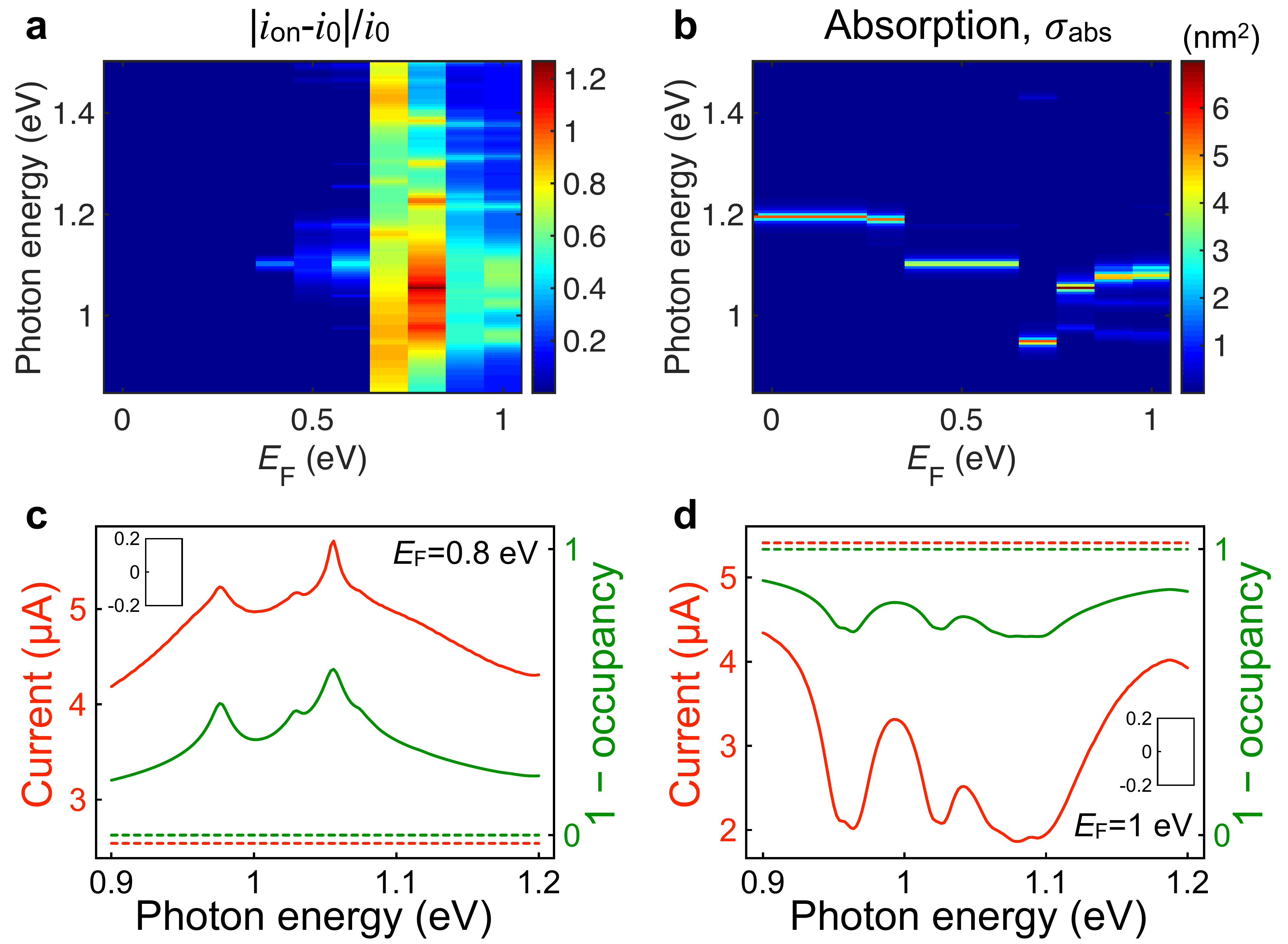}
\par\end{centering}
\caption{{\bf Plasmometer based on a GHQD junction.} ({\bf a}) Relative variation of the electric current through the device of Fig.\ \ref{Fig1}a (color scale) as a function of both incident photon energy and Fermi energy for a fixed DC bias voltage of 0.1 V and a light fluence of 30\,mJ$/$m$^{2}$. ({\bf b}) Absorption cross-section of the GHQD under the same condition as in {\bf a}. ({\bf c,d}) Photon-energy dependence of the predicted current (left scale) and depletion of states near the Fermi level ($=1-$occupancy, right scale) with/without external illumination (solid/dashed curves). The Fermi energy is $\EF=0.8\,$eV in {\bf c} and $\EF=1\,$eV in {\bf d}. The positions of relevant energy levels relative to $\EF$ are shown as insets to {\bf c} and {\bf d} ({\it i.e.}, the plotted occupancy is the average over those levels).}
\label{Fig3}
\end{figure}

Tunability is one of the most salient properties of graphene plasmons.  The spectral position of the plasmon resonance supported by our proposed device (Fig.\ \ref{Fig1}a) can be varied over a wide spectral range {\it via} chemical doping or electrical gating, as shown in Fig.\ \ref{Fig3}b. Incidentally, Fig.\ \ref{FigS3} shows a smoother dependence of the excitation features on doping charge when the latter is changed as a continuous variable; the lower-right feature in Figs.\ \ref{Fig3}b and \ref{FigS3} is then identified as a plasmon with a characteristic increase of frequency with doping). It is reassuring to observe that the corresponding electrical signal (Fig.\ \ref{Fig3}a) shows a similar dependence on both the doping level and the photon energy as the optical absorption of the GHQD (Fig.\ \ref{Fig3}b), further supporting the ability of the nanojuction to serve as a plasmometer ({\it i.e.}, as a nanoscale spectrometer from the point of view of the external illumination). We note that the stronger spectral variations of electrical current are observed at higher Fermi energies due to the availability of a denser set of electronic states that are influenced by optical heating. Remarkably, the variation of the electric current follows rather closely the evolution of the occupancy of electron states in the vicinity of the Fermi level ({\it i.e.}, the creation or blocking of additional channels of electrical conductance), with strong features emerging near the plasmon resonances (see Fig.\ \ref{Fig3}c,d, where the insets show the relevant electronic states relative to the Fermi level). More precisely, the current is roughly proportional to one minus the occupancy, leading to either enhancement (Fig.\ \ref{Fig3}c) or depletion (Fig.\ \ref{Fig3}d) of the electrical signal, depending on the specific distribution of the relevant electronic states around the Fermi level.

\begin{figure}
\begin{centering}
\includegraphics[width=0.6\textwidth]{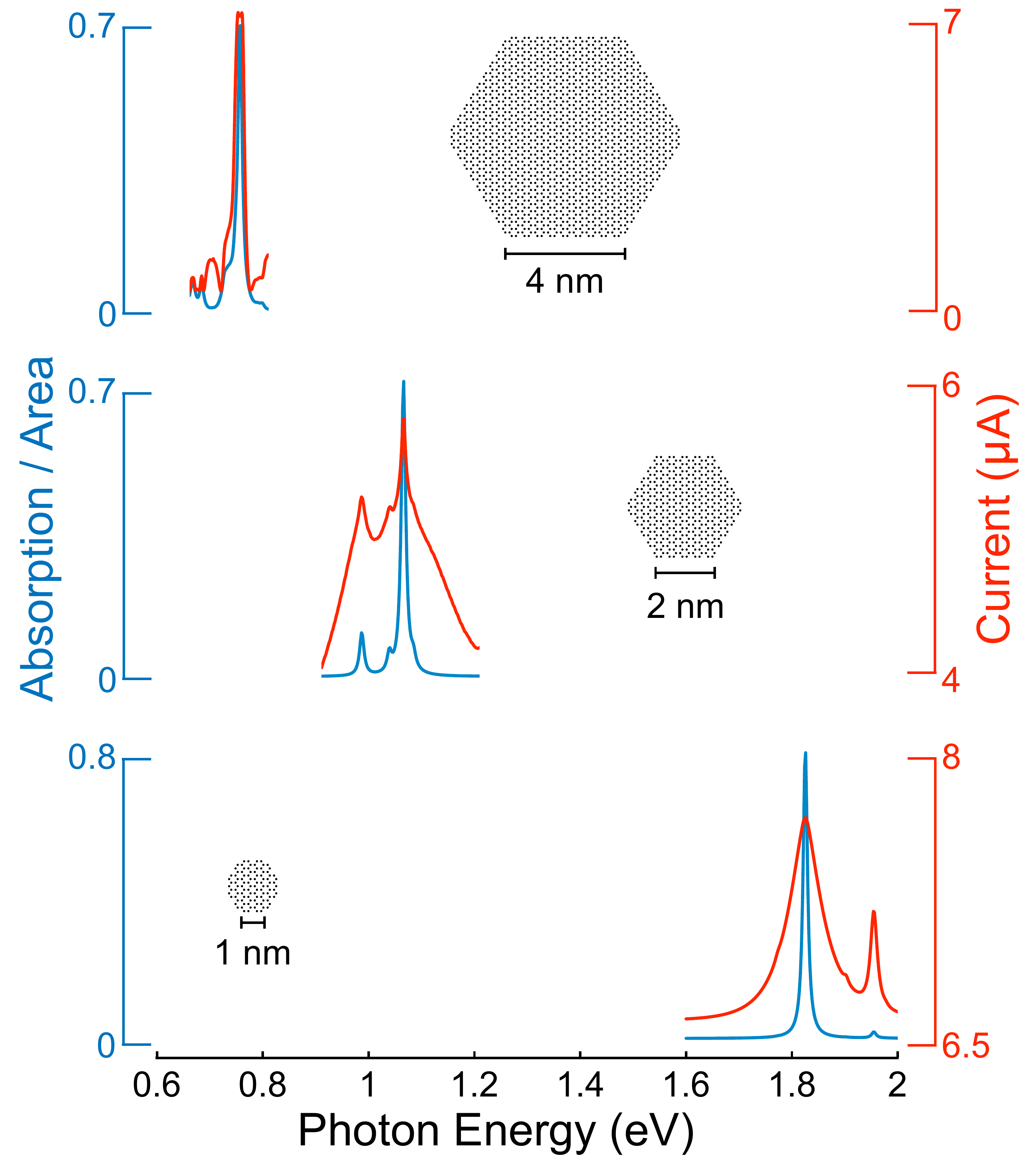}
\par\end{centering}
\caption{{\bf Size dependence of the plasmometer performance.} We present the optical absorption cross-section (left scales) and the electric current through the plasmometer device (right scales) as a function of incident photon energy for GHQDs of different side lengths (4, 2, and 1\,nm, from top to bottom).}
\label{Fig4}
\end{figure}

An alternative way of shifting the plasmon energy consists in changing the size of the GHQD. A series of nanoscale devices similar to the one considered above could then act simultaneously to detect different light wavelengths. As we show in Fig.\ \ref{Fig4}, when the size of the GHQD decreases, the main plasmon resonance moves to higher photon energies, thus covering the near-infrared spectral region. Importantly, the resonance features in the electric-current spectra also shift to higher photon energies (Fig.\ \ref{Fig4}, right scale) and actually coincide with the resonance peaks in the absorption cross-section (left scale). This further corroborates the excellent performance of the proposed nanojunction as a spectral plasmometer device.

\section{Concluding remarks}

In summary, we have shown through state-of-the-art quantum-mechanical simulations that a compact graphene nanojunction constitutes a device capable of yielding the number of plasmons excited in a central graphene island with a sensitivity down to the single plasmon level. The conductivity of the junction is shown to be severely modified by the presence of excited plasmons, thus permitting a direct readout of the number of such excitations through a measurement of the electric current for a given bias voltage applied between the gates on either side of the junction. This strong dependence of a nanographene quantum-dot conductivity on its optical excitation state is inherited from the peculiar band structure of this material, through essentially the same mechanism that endows it with a strong electro-optical tunability: small changes in the electronic structure are amplified through the participation of a large number of conduction electrons in the optical response \cite{paper235}, but also in the electric transport properties. We argue that this plasmometer functionality is the basis for the application of the device as a spectrometer, which can achieve spectral selectivity by gate-controlling the plasmon frequency, or alternatively, by arraying several devices with different junction sizes.

We have presented results assuming a moderate value of the plasmon lifetime (66\,fs). In our model, this parameter affects the absorption cross-section, with longer lifetimes expected to result in larger absorption for fixed light pulse fluence ---an effect that only adds a scale factor depending on the light intensity. Additionally, we remark that pulse duration, plasmon lifetime, and electron thermalization span a time interval $\sim$10's\,fs, which is small compared with the time over which the electron gas remains at an elevated temperature (100's\,fs) before inelastic relaxation builds up. It is during this period of high-electron temperature when the a modified current should be observed. This establishes a clear temporal hierarchy for the relevant processes under consideration, and although we are neglecting current variations during the transient period of initial plasmon excitation and nonequilibrium electron distribution, they are clearly an exciting topic for future studies outside the scope of the present work.

It should be noted that leakage from the island into the leads can occur when its electron distribution is at an elevated temperature, thus adding a net charge to the island that needs to be neutralized by additional opposite charge flowing into the island. Essentially, this constitutes a current that flows through the structure, which is precisely what the Landauer formalism allows us to calculate. A relevant question arises, to what extend is this current contributing to reduce the electron temperature in the island ({\it i.e.}, through inelastic transmission into the leads, for example with electron-electron scattering resulting in higher-energy electrons flowing to the leads at the expense of lower-energy electrons moving into the island). Given the small size of the contacts, we neglect these processes under the assumption that they only produce a small correction.

An additional consideration relates to the linear scaling here assumed for the number of excited plasmons as a function of the absorption cross-section. In this respect, saturable absorption can introduce nonlinearity, although the characteristic saturation fluence is $\sim10$\,J$/$m$^2$ for 100\,fs pulses \cite{paper247}, which is large compared with the values $<50$\,mJ$/$m$^2$ here considered. Perhaps more importantly, plasmon decay might depend on the presence of nonthermal electrons produced by previously deceased plasmons, which can create Pauli blocking and also increase Landau damping, so its net effect is difficult to assess. This modification of the plasmon lifetime affects the transient period during which the plasmon stays alive, and therefore, we neglect it in this work. Nonetheless, this line of thought could give rise to unexplored ultrafast physics phenomena that are worth investigating in future work.

In our design, we have considered the central island and the electrical contacts to be all made of graphene, which seems to be a logical choice starting from an extended graphene sheet, from which the junction structure could be patterned using nanolithography. Although we present calculations only for nanometer-sized junctions, which display plasmons in the near infrared, the proposed concept can be scaled up in size to junctions that are several tens of nanometers in size, which should allow us to access the mid-infrared spectral region, in which currently available spectrometers generally have poor efficiencies. Additionally, the plasmometer could be used to measure the number of plasmons produced by means other than external illumination, such as fluorescence from nearby optical emitters ({\it e.g.}, labels for sensing) or free electron beams. Our plasmometer design therefore holds great potential for on-chip nanophotonic devices comprising photodetectors, spectrometers, and sensors, as well as for nanoscale quantum devices that benefit from its ability to resolve single plasmons.

\begin{figure}
\begin{centering}
\includegraphics[width=0.4\textwidth]{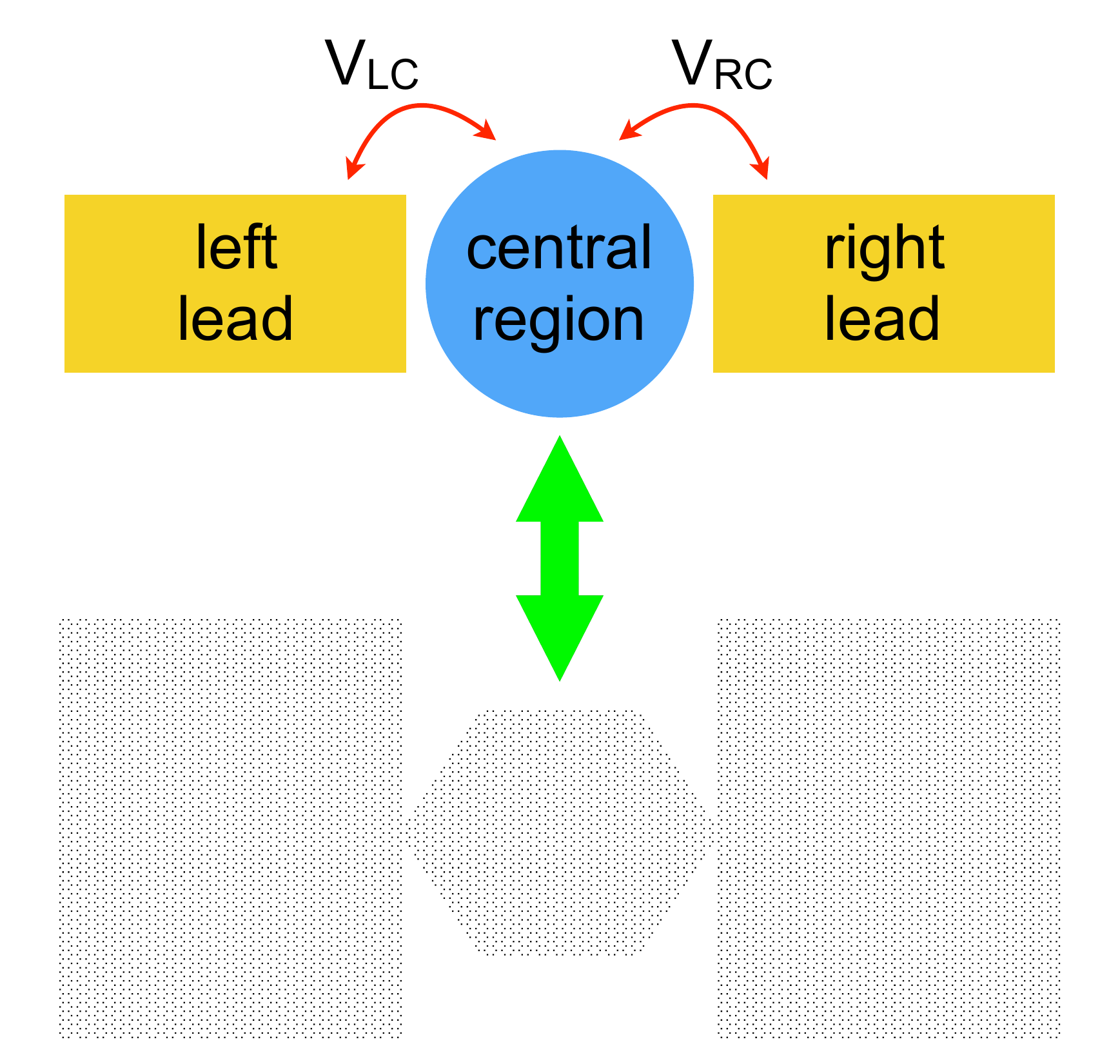}
\par\end{centering}
\caption{Schematics of our proposed graphene hexagon quantum dot (GHQD) junction, consisting of a finite central region coupled to two semi-infinite leads {\it via} the potentials $V_{\rm LC}$ and $V_{\rm RC}$.}
\label{Fig5}
\end{figure}

\appendix

\section{Landauer's formalism}

It is natural to divide our system into three parts as shown in Fig.\ \ref{Fig5}: a central region (Hamiltonian $H_{\rm C}$, see below) and the left and right semi-infinite graphene sheets ({\it i.e.}, the leads of Hamiltonians $H_{\rm L}$ and $H_{\rm R}$, respectively). The couplings between the central region and the leads are described by potentials $V_{\rm LC}$ and $V_{\rm RC}$, respectively. With these definitions, the total Hamiltonian $H=H_{\rm C}+H_{\rm L}+H_{\rm R}+V_{\rm LC}+V_{\rm RC}+V_{\rm LC}^\dagger+V_{\rm RC}^\dagger$ allows us to project the Schr\"{o}dinger equation for an electron energy $E$ into single-particle wave functions associated with each of the three regions, $|\phi_{\rm L}\rangle$,  $|\phi_{\rm C}\rangle$, and $|\phi_{\rm R}\rangle$. We find \cite{D97,V08}
\begin{equation}
\left(E-H_{\rm L}\right)|\phi_{\rm L}\rangle=V_{\rm LC}|\phi_{\rm C}\rangle,\label{eq:eq1}
\end{equation}
\begin{equation}
V_{\rm LC}^\dagger|\phi_{\rm L}\rangle+H_{\rm C}|\phi_{\rm C}\rangle+V_{\rm RC}^\dagger|\phi_{\rm R}\rangle=E|\phi_{\rm C}\rangle,\label{eq:eq2}
\end{equation}
\begin{equation}
\left(E-H_{\rm R}\right)|\phi_{\rm R}\rangle=V_{\rm RC}|\phi_{\rm C}\rangle.\label{eq:eq3}
\end{equation}
We now rewrite Eqs.\ (\ref{eq:eq1}) and (\ref{eq:eq3}) as
\begin{align}
|\phi_{\rm L}\rangle=G_{\rm L}^{\pm}V_{\rm LC}|\phi_{\rm C}\rangle,\label{eq:GL}
\end{align}
\begin{equation}
|\phi_{\rm R}\rangle=G_{\rm R}^{\pm}V_{\rm RC}|\phi_{\rm C}\rangle,\label{eq:eq3a}
\end{equation}
where $G_{\rm L,R}^{\pm}=\left(E\pm\mathrm{i}\epsilon-H_{\rm L,R}\right)^{-1}$ are the retarded ($+$) and advanced ($-$) Green functions of the left ($L$) and right ($R$) leads, while $\epsilon\rightarrow0^+$ is a positive infinitesimal. For simplicity, we assume the Green functions to be diagonal and given by $G_{\rm L,R}^{\pm}=-i\pi$ ({\it i.e.}, the local density of states at the leads \cite{TDH98}). Substituting Eqs.\ (\ref{eq:GL}) and (\ref{eq:eq3a}) into Eq.\ (\ref{eq:eq2}), we find
\begin{equation}
\left[E-H_{\rm C}-V_{\rm LC}^\dagger G_{\rm L}^{\pm}V_{\rm LC}-V_{\rm RC}^\dagger G_{\rm R}^{\pm}V_{\rm RC}\right]|\phi_{\rm C}\rangle=0.\label{eq:eqFinal}
\end{equation}
It is then convenient to define self-energy operators $\Sigma_{\rm L}^{\pm}=V_{\rm LC}^\dagger G_{\rm L}^{\pm}V_{\rm LC}$ and $\Sigma_{\rm R}^{\pm}=V_{\rm RC}^\dagger G_{\rm R}^{\pm}V_{\rm RC}$, in terms of which the retarded ($+$) and advanced ($-$) Green functions of the central region in the presence of the leads ({\it i.e.}, the Green functions of Eq.\ (\ref{eq:eqFinal})) reduce to
\begin{equation}
G^{\pm}=\left(E-H_{\rm C}-\Sigma_{\rm L}^{\pm}-\Sigma_{\rm R}^{\pm}\right)^{-1}.\nonumber
\end{equation}
From here, we calculate the transmission function \cite{D97,V08} (from left to right) as
\begin{equation}
T(E,V)=\mathrm{Tr}\left\{\Gamma_{\rm L}G^{+}\Gamma_{\rm R}G^{-}\right\},\nonumber
\end{equation}
where $\Gamma_{\rm L,R}=\mathrm{i}\left(\Sigma_{\rm L,R}^{+}-\Sigma_{\rm L,R}^{-}\right)$. Finally, the Landauer formula \cite{L1957,D97,V08,TDH98} allows us to relate the current to the voltage (\textit{I}-\textit{V} curves) as
\noindent 
\begin{equation}
I=\frac{e}{\pi\hbar}\int_{-\infty}^{+\infty}dE\;\;T(E,V)\left[f_{T_{\rm c}}\left(E-\mu_{\rm L}\right)-f_{T_{\rm c}}\left(E-\mu_{\rm R}\right)\right],\nonumber
\end{equation}
where $\mu_{\mathrm{L}}=E_{\mathrm{F}}-eV/2$ and $\mu_{\mathrm{R}}=E_{\mathrm{F}}+eV/2$ are the electrochemical potentials in the two contacts, which are assumed to remain in local equilibrium,
\begin{equation}
f_T(E-\mu)=\frac{1}{1+\ee^{(E-\mu)/k_{\mathrm{B}}T}}
\nonumber
\end{equation}
is the Fermi-Dirac distribution at an electron temperature $T$, and $k_{\mathrm{B}}$ is the Boltzmann constant. In particular, we assume the temperature at the contacts $T=T_{\rm c}=30\,$K to remain constant throughout this work.

\section{GHQD Hamiltonian}

We describe the electronic structure of the central island through a nearest-neighbors tight-binding Hamiltonian $H_{\rm TB}$ incorporating one state per carbon atom and a hopping energy of $t=2.8\,$eV \cite{W1947,CGP09}. The GHQD Hamiltonian
\begin{equation}
H_{\rm C}=H_{\rm TB}+H_{\mathrm{Hartree}}+V_{\mathrm{bias}}\nonumber
\end{equation}
additionally includes the potential due to the DC biased contacts $V_{\mathrm{bias}}$ and the self-consistent Hartree interaction among conduction electrons $H_{\mathrm{Hartree}}$. For simplicity, we assume $V_{\mathrm{bias}}$ to vary linearly across the gap separating the two gates. The Hartree interaction is diagonal in the cabon-site representation \cite{paper183,paper247}:
\begin{equation}
H_{\mathrm{Hartree},ll'}=2\delta_{ll'}\sum_{jl^{''}}v_{ll^{''}}f_{T_{\rm e}}(E_j-\mu)\left|a_{jl^{''}}\right|^{2}\label{eq:new Hartree}
\end{equation}
where $l$, $l'$, and $l''$ run over carbon sites, $j$ runs over one-electron states of energy $E_j$ and wave function coefficients $a_{jl}$, $v_{ll^{''}}$ describes the Coulomb interaction between sites $l$ and $l''$, $f(E_j-\mu)$ is the Fermi-Dirac distritution (see above), $\mu$ is the chemical potential, and $T_{\rm e}$ is the electron temperature in the island. The leading factor of 2 in Eq.\ (\ref{eq:new Hartree}) stems from spin degeneracy. The electronic states of the system are then calculated following a series of self-consistent iterations until convergence is achieved in the Hartree term.

\noindent \textbf{Electron temperature.} The Fermi-Dirac distribution depends on both the electron temperature $T_{\rm e}$ in the GHQD and the $T_{\rm e}$-dependent chemical potential $\mu$. The latter is determined from the condition that the total number of electrons is conserved. We perform calculations without optical pumping by assuming $T_{\rm e}=30\,$K, whereas \textit{I}-\textit{V} curves after pulse irradiation are obtained assuming a temperature determined from the optically absorbed energy. More precisely, after excitation and subsequent decay of plasmons in the GHQD, the system is thermalized within 10's fs, reaching an electron temperature \cite{paper235}
\begin{equation}
T_{\mathrm{e}}=\left[\frac{F_{\mathrm{in}}\sigma_{\mathrm{abs}}\left(\hbar v_{\mathrm{F}}\right)^{2}}{2.3Ak_{\mathrm{B}}^{3}}\right]^{\frac{1}{3}},\label{Te}
\end{equation}
where $\sigma_{\mathrm{abs}}$ is the absorption cross section (see below), $F_{\mathrm{in}}$ is the incident light fluence, $v_{\mathrm{F}}\approx10^{6}\,$m$/$s is the Fermi velocity, and $A$ is the graphene area. Incidentally, the 2.3 factor in the denominator of Eq.\ (\ref{Te}) is corrected with a factor of 2 that was missing in Ref.\ \cite{paper235}.

\noindent \textbf{Optical response.} We simulate the optical response of graphene by using the random-phase approximation to calculate the noninteracting susceptibility \cite{PN1966,paper183,paper247}. The external perturbation from the incident light is introduced through a potential $\phi^{\mathrm{ext}}(\rb)=-\rb\cdot\Eb^{\mathrm{ext}}$, where $\Eb^{\mathrm{ext}}$ is the light electric field. This allows us to compute the absorption cross-section $\sigma_{\mathrm{abs}}$ as described elsewhere \cite{paper183}.

\begin{figure}
\begin{centering}
\includegraphics[width=0.6\textwidth]{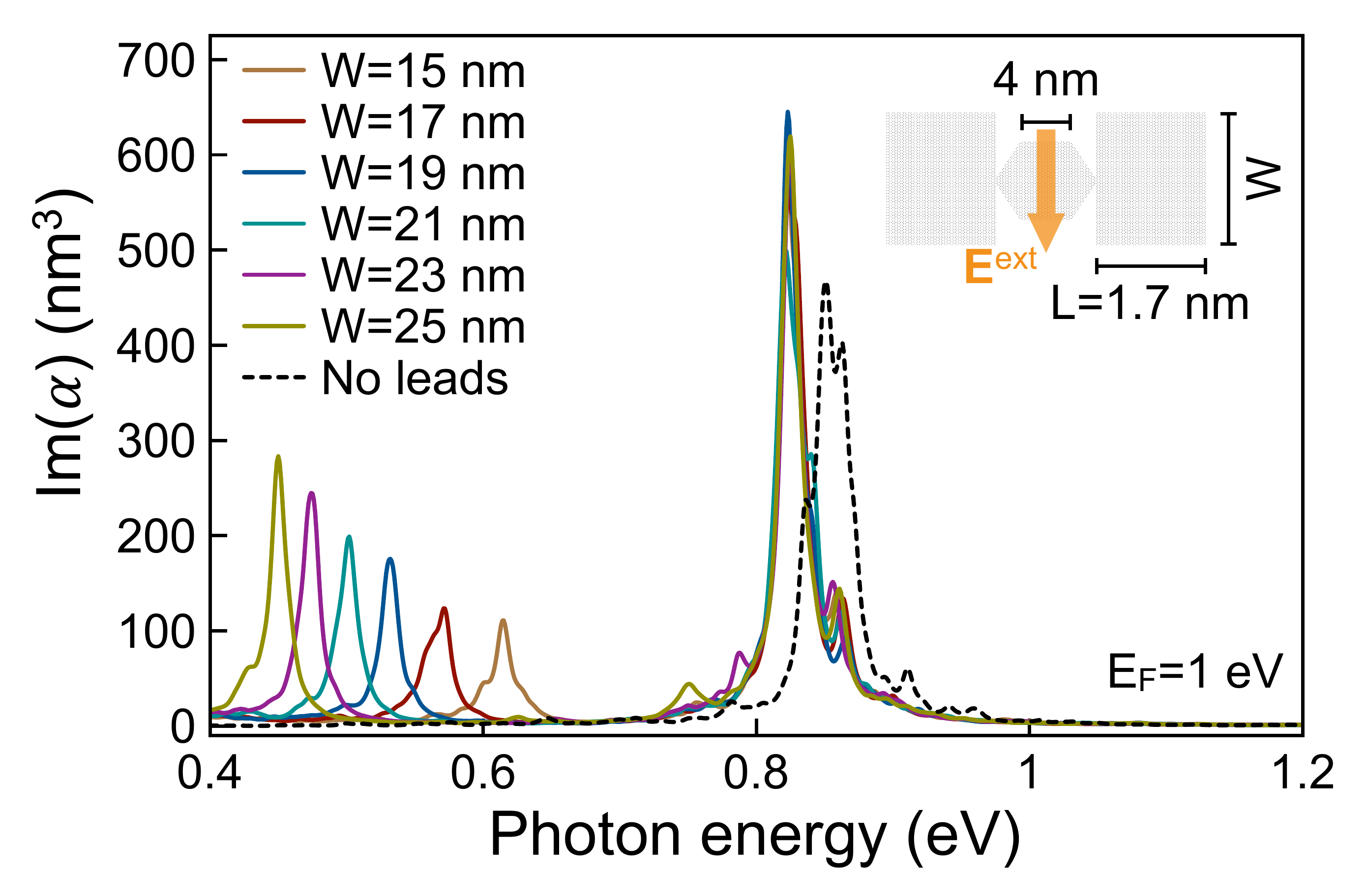}
\par\end{centering}
\caption{Polarizability spectra for nanojunctions consisting of a doped GHQD (4\,nm side length in all cases) and finite leads of different dimensions (see inset and legend). The Fermi energy is 1\,eV in all cases.}
\label{Fig6}
\end{figure}

Figure\ \ref{Fig6} shows that the optical response of an isolated GHQD (dashed curve) in slightly broadened and blue shifted with respect to the response of the island in the presence of the leads. These calculations are performed for leads of finite width, which produce additional low-energy features in the spectral response, although they disappear in the limit of large widths. Through further calculations (not shown), we corroborate that these conclusions are maintained when considering GHQDs of different size and Fermi energies within the ranges considered in this work. For simplicity, we thus assume that the GHQD is isolated when calculating its optical response in order to obtain the optical absorption and from here the electron temperature according to Eq.\ (\ref{Te}).

\begin{figure}
\begin{centering}
\includegraphics[width=0.6\textwidth]{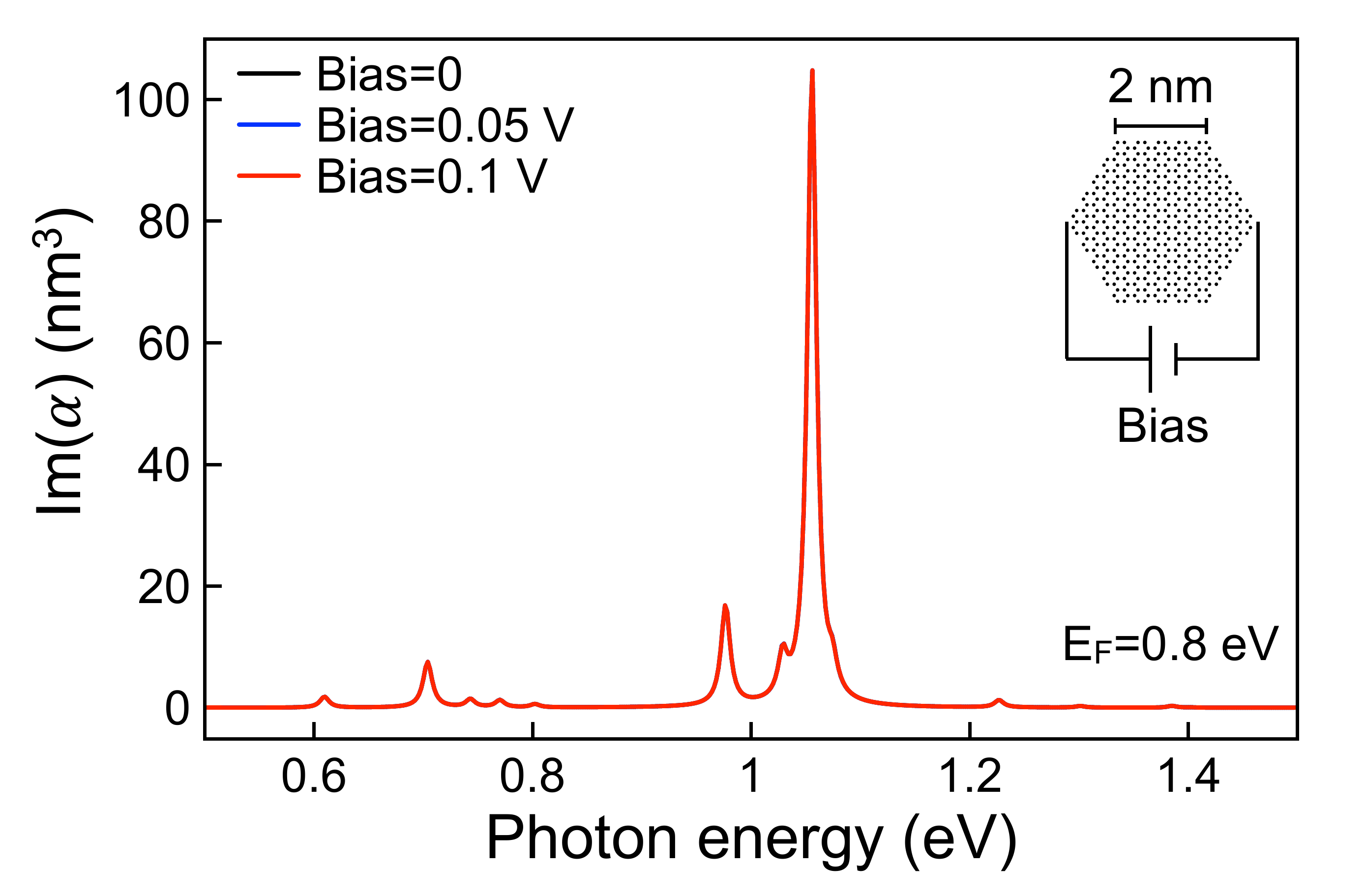}
\par\end{centering}
\caption{Polarizability spectra of a single GHQD under different
external DC bias voltages, essentially overlapping in a single visible curve.}
\label{Fig7}
\end{figure}

Additionally, the response of the GHQD does not change significantly when applying a small lateral DC bias within the voltage range (0-0.1\,V) here considered (see Fig.\ \ref{Fig7}). Consequently, we also neglect the effect of the lateral bias on the optical response of the GHQDs throughout the rest of this work.

\noindent \textbf{Electronic DOS in the graphene leads.} We have calculated the DOS at the edge of a doped semi-infinite graphene sheet by direct summation of tight-binding electron states in ribbons of increasing width. We find the result to be nearly indistinguishable from the DOS for infinite graphene, which in the range of electron energies $E$ under consideration ($-t\leqslant E\leqslant t$) reduces to the analytical expression \cite{CGP09}
\begin{equation}
	D\left(x\right)=\frac{8}{\pi^{2}}x^{2}\frac{1}{\sqrt{4(1+x)^{2}-(x^{2}-1)^{2}}}\mathrm{\mathbf{F}}\left(\frac{\pi}{2},4\sqrt{\frac{x}{4(1+x)^{2}-(x^{2}-1)^{2}}}\right),\label{eq:DOS}
\end{equation}
where $x=|E/t|$ and $\mathbf{F}\left(\pi/2,y\right)$ is the complete elliptic integral of the first kind.

\noindent \textbf{Number of plasmons and plasmon-to-charge conversion efficiency.} The average number of plasmons excited in the nanojunction by an incident pulse of fluence $F_{\mathrm{in}}$ is given by
\begin{equation}
	n_{\rm p}=\frac{F_{\mathrm{in}}\sigma_{\mathrm{abs}}}{\hbar\omega_{\mathrm{p}}},\label{eq:number}
\end{equation}
where $\hbar\omega_{\mathrm{p}}$ is the plasmon energy and $\sigma_{\mathrm{abs}}$ is the absorption cross-section. We further define the plasmon-to-charge conversion efficiency as the ratio\cite{FKC09,GCB15}
\begin{equation}
	\eta=\frac{|i_{\mathrm{on}}-i_{0}|}{e}\frac{\hbar\omega_{\mathrm{p}}\Delta t_{\rm eff}}{F_{\mathrm{in}}\sigma_{\mathrm{abs}}},
\label{eq:efficiency}
\end{equation}
which gives the additional number of electrons circulating through the junction as a result of plasmon excitations, normalized to the number of excited plasmons ({\it i.e.}, each excited plasmon produces the circulation of $\eta$ additional electrons through the junction). Here, $i_{\mathrm{on}}$ and $i_{0}$ are the currents under resonant illumination and in the absence of illumination, respectively, while $\Delta t_{\rm eff}$ is an effective time interval during which the electron gas remains at an elevated temperature after pulse irradiation. We take $\Delta t_{\rm eff}=200\,$fs as a conservative estimate \cite{DSC08,WOP11}.

\begin{figure}
\begin{centering}
\includegraphics[width=0.7\textwidth]{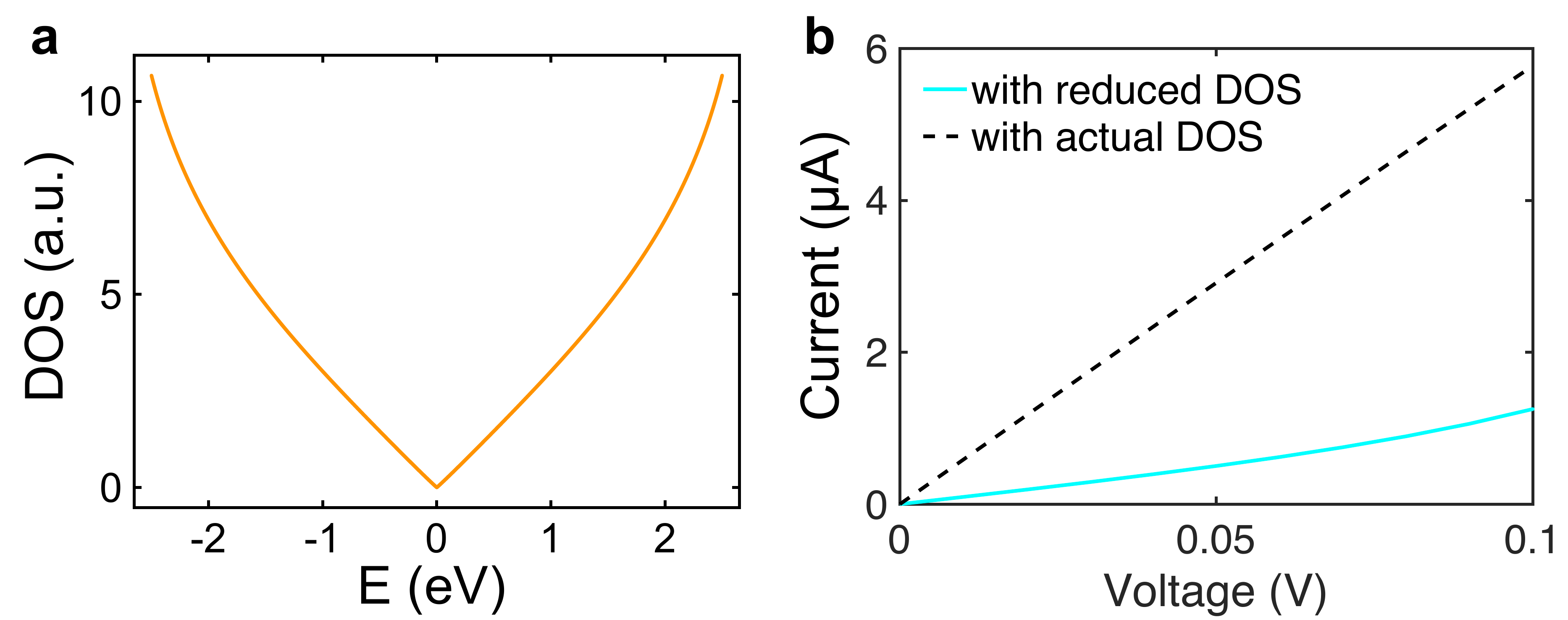}
\par\end{centering}
\caption{(a) DOS of graphene leads. (b) Simulated currents as a function of bias voltages with the same on-resonance conditions and central GHQD as in Fig.\ 1. The cyan solid curve represents the \textit{I}-\textit{V} response assuming a DOS for the graphene leads ten times smaller than that in Fig.\ 1. The black dashed curve is taken from Fig.\ 1c and shown as a reference.}
\label{FigS1}
\end{figure}

\begin{figure}
\begin{centering}
\includegraphics[width=0.5\textwidth]{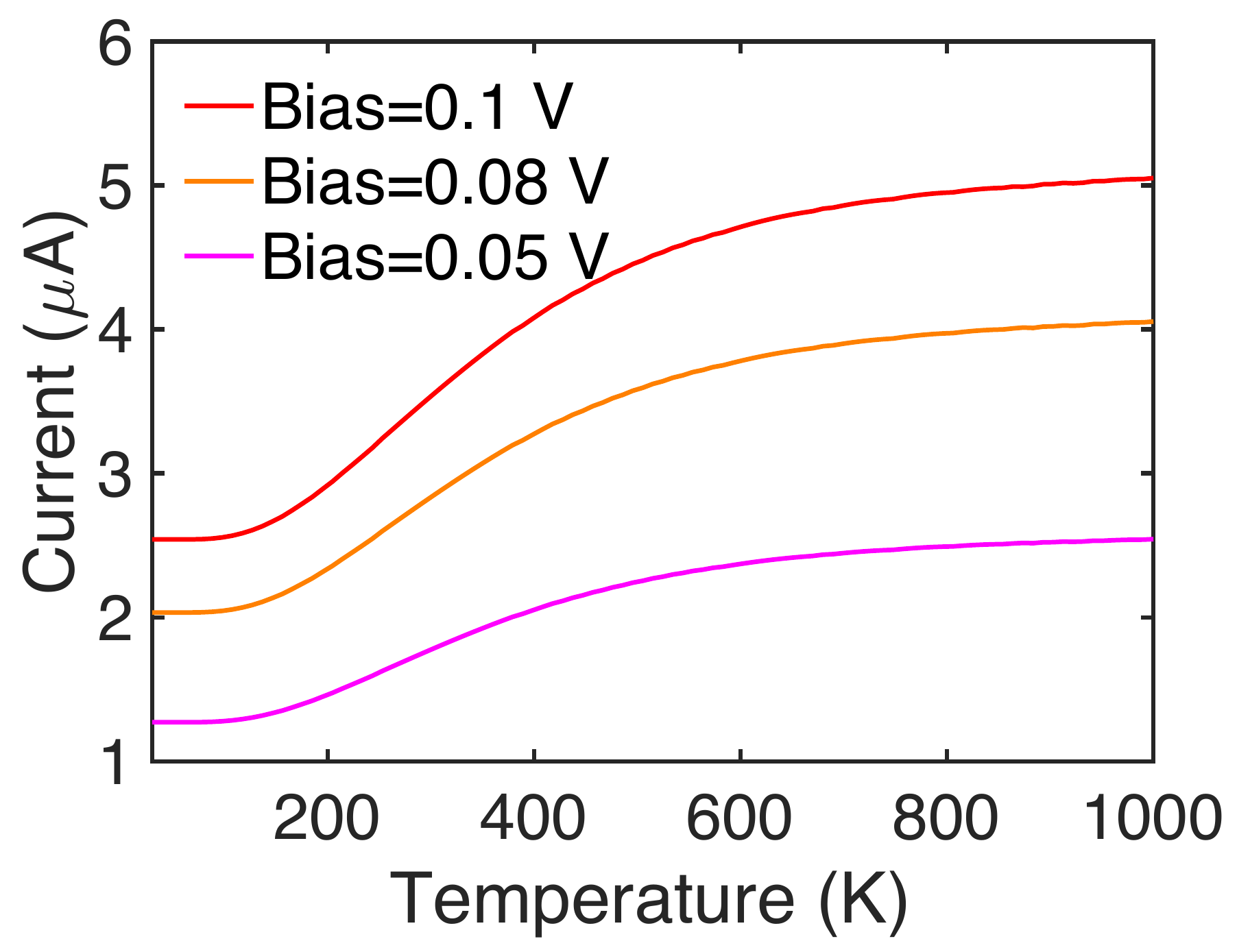}
\par\end{centering}
\caption{Predicted current under the same conditions as in Fig.\ 2a for a range of electron temperatures corresponding to fluences in the 0-0.7\,mJ$/$m$^2$ region, according to Eq.\ (8).}
\label{FigS2}
\end{figure}

\begin{figure}
\begin{centering}
\includegraphics[width=0.5\textwidth]{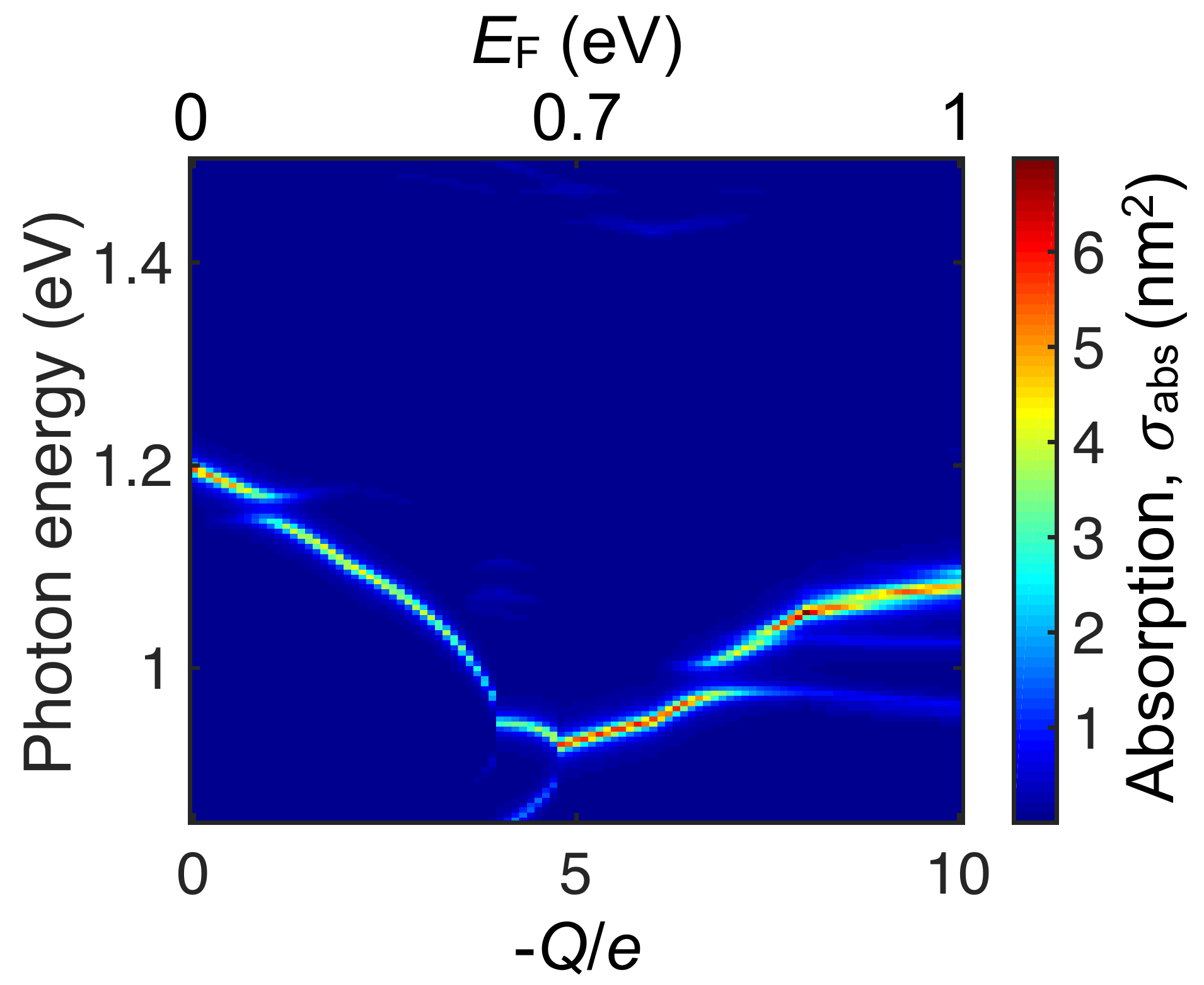}
\par\end{centering}
\caption{Same as Fig.\ 3b, plotted as a function of the doping charge $Q$, which is taken to be a continuous variable ({\it i.e.}, fractional doping charges are considered in order to visualize the smooth evolution of the absorption cross-section as a function of this parameter).}
\label{FigS3}
\end{figure}

\section{Density of States (DOS) of Graphene Leads and DOS Dependence of the
\textit{I}-\textit{V} Curves}

In Fig.\ \ref{FigS1}a, we show the DOS per carbon site at the graphene leads (a part of it is also shown in Fig.\ 1d), which admits an analytical expression \cite{CGP09} (see Eq.\ (9)). When this DOS is artificially reduced, the current also decreases, as shown in Fig.\ \ref{FigS1}b.

\section{Low-Fluence Limit of the Current under the Conditions of Figure\ 2a}

In Fig.\ \ref{FigS2}, we show the current obtained under the same conditions as in Fig.\ 2a for electron temperatures in the 30-1000\,K range, corresponding to fluences in the 0-0.7\,mJ$/$m$^2$ ranges according to Eq.\ (8).

\section*{Absorption Cross-Section for Continuously Varying Doping under the Conditions of Figure\ 3b}

Figure\ \ref{FigS3} is a recalculation of Fig.\ 3b considering a continously varying doping charge. Fractional doping charges could correspond to a situation in which electrons or holes tunnel into the island from a neiboring gate, so that their wave functions are shared between the island and the gate.

\section{Acknowledgments}

This work has been supported in part by the Spanish MINECO (MAT2014-59096-P and SEV2015-0522), AGAUR (2014 SGR 1400), Fundaci\'o Privada Cellex, and the European Commission (Graphene Flagship CNECT-ICT-604391 and FP7-ICT-2013-613024-GRASP).


\end{document}